\documentclass[seceq]{ptptex}
\usepackage{wrapft}
\usepackage{graphicx}
\newcommand{\bra}[1]{\langle {#1} |}     %%
\newcommand{\ket}[1]{| {#1} \rangle}     %%
\newcommand{\wtilde}[1]{\widetilde{#1}} %%

\newcommand{\ovl}[1]{\overline{#1}}
\newcommand{\lvect}[1]{\overleftarrow {#1}}

\def\beq{\begin{eqnarray}}
\def\eeq{\end{eqnarray}}
\def\bsub{\begin{subequations}}
\def\esub{\end{subequations}}
\def\b{\begin{equation}}
\title{%        %You can use \\ for explicit line-break
Time-Dependent Variational Approach to the Non-Abelian\\
Pure Gauge Theory
}
\subtitle{
Its Application to Evaluation of the 
Shear Viscosity of\\
Quantum Gluonic Matter 
}    %use this when you want a subtitle

\author{%       %Use \sc for the family name
Yasuhiko {\sc Tsue}${}^{1}$, Tong-Gyu {\sc Lee}${}^2$ and 
Hiroshi {\sc Ishii}${}^{1}$
\footnote{Present address : Suzuki Motor Corporation}
}

\inst{%         %Affiliation, neglected when [addenda] or [errata]
${}^1$Physics Division, Faculty of Science, Kochi University, Kochi 780-8520, 
Japan\\
${}^2$Graduate School of Integrated Arts and Sciences, Kochi University, 
Kochi 780-8520, Japan
}

\recdate{%      %Editorial Office will fill in this.
\today
}

\abst{%       %this abstract is neglected when [addenda] or [errata]
The time-dependent variational approach to the pure Yang-Mills gauge 
theory, especially a color $su(3)$ gauge theory, is formulated in the 
functional Schr\"odinger picture with a Gaussian wave functional 
approximation. 
The equations of motion for the quantum gauge fields are formulated 
in the Liouville-von Neumann form. 
This variational approach is applied in order to derive the transport 
coefficients, such as the shear viscosity, for the pure gluonic matter 
by using the linear response theory. As a result, the contribution 
to the shear viscosity from the quantum gluons is zero up to the 
lowest order of the coupling $g$ in the quantum gluonic matter. 
}

\begin{document}

\maketitle

\section{Introduction}

One of recent interests for the quark and gluon physics, governed by 
the quantum chromodynamics (QCD), is to investigate properties of the 
quark-gluon plasma (QGP) and/or the quark-gluon matter. 
In the recent progress of the Relativistic Heavy Ion Collider (RHIC) 
experiments, it is said that the QGP may be not free gas 
but the strongly interacting quark-gluon matter.\cite{QM06}
The matter composed of quarks and gluons seems to reveal 
the properties of the 
liquid, not gas, like the perfect liquid. 
This conjecture is derived by comaparing the obtained experimental data 
with the phenomenogical analysis by using the hydrodynamical simulation 
with rather small shear viscosity, which leads to the 
near perfect liquid. 
The small shear viscosity is also near the lower bound which is 
conjectured in the AdS/CFT correspondence.\cite{AdS}

Many works to understand properties of the gluonic matter were 
performed recently,\cite{Greiner,Meyer,Yaffe} 
while the transport coefficients, especially 
the shear viscosity for pure gluonic matter, were evaluated up to 
the lowest order of the QCD coupling constant $g$ in the early 
study.\cite{Kajantie}
Namely, the shear viscosity $\eta_C$ for the gluonic matter at temperature 
$T$ can be 
expressed as\cite{Kajantie,Yaffe}
\beq\label{1-1}
& &\eta_C = d_f \frac{T^3}{g^4\log(1/g^2)} \ ,
\eeq
up to the lowest order of $g$. 
Here, $d_f$ is numerically determined constant. 
For the quark matter, the shear viscosity is also evaluated, for example, 
in the Nambu-Jona-Lasinio (NJL) model\cite{NJL} 
by using the linear response theory,\cite{Kubo,Mori} 
in which a rather small shear viscosity 
is derived.\cite{Iwasaki} 
The small shear viscosity leads to the short mean free path in general. 
Thus, it may be shown that the constituents, namely quarks and gluons, 
of the matter under consideration, are strongly correlated. 

However, as was shown by Ref.\citen{Asakawa}, 
the anomalous contribution to the shear viscosity 
in the turbulent plasma fields 
gives the small shear viscosity even in 
the weak coupling QCD in which the quarks and gluons are weakly 
correlated. 
Namely, the shear viscosity $\eta$ can be expressed by including 
the anomalous contribution $\eta_A$ as\cite{Asakawa} 
\begin{equation}\label{1-2}
\eta^{-1}=\eta_C^{-1}+\eta_A^{-1}\ .
\end{equation}
If the anomalous viscosity $\eta_A$ is small, the total shear 
viscosity $\eta$ becomes to small even if the usual shear viscosity 
$\eta_C$ is large for the small coupling constant $g$ in Eq.(\ref{1-1}). 
Thus, the small shear viscosity does not always lead to the strong 
coupling QCD, namely strongly correlated quark-gluon matter.

In this paper, 
thus, we consider the pure gluonic matter as the weak coupling system. 
we are concentrated our interest to calculating the shear 
viscosity for pure quantum gluonic matter without quarks by using 
the linear response theory. 
In this paper, the quantum gluon means the quantum fluctuation part 
around the mean field which leads to the Eq.(\ref{1-1}). 
Thus, the shear viscosity of the quantum gluonic matter gives 
the contribution of the next and 
higher order of $g$ in comparison with Eq.(\ref{1-1}). 
One of purposes in this paper is to investigate the shear viscosity 
under small QCD coupling $g$ for the quantum gluonic matter. 
It is important to investigate the contribution to the shear viscosity 
from the quantum gluonic fields. 
The reason is as follows: 
If there is a contribution to the shear viscosity of the order of $g^0$, 
the finite value of the shear viscosity is remained even if the coupling 
$g$ is small.

To deal with the quantum gluons and to investigate the dynamics of the 
quantum gluons, the time-dependent variational method with the 
Gaussian functional as a trial wave functional in the functional 
Schr\"odinger picture may gives a useful tool.\cite{Dominique} 
The reason why is that the mean fields and the quantum fluctuations 
around them can be treated on an equal footing and the 
higher order contributions for $g$ are automatically included 
because certain kinds of the Feynman diagrams are taken into account 
in this variational approach. 
In this variational approach, the equations of motion 
for the mean fields and the fluctuation modes around them are derived 
in a self-consistent manner. 
Especially, the equations of motion for the quantum gluon fields 
are formulated in a form of the Liouville-von Neumann equations.

Another merit to use the time-dependent variational method for the pure 
gluonic matter is that the expectation values for various 
field operators and their products can easily be calculated 
because the state or the wave functional is prepared in the process of the 
variational calculations. 
When the transport coefficients such as the shear viscosity are calculated 
by using the linear response theory, the expectation values or thermal 
averages for the various operators such as the energy-momentum 
tensor operator are necessary. 
Thus, the variational approach may be suitable and give a practical 
method to calculate the transport coefficients.

This paper is organized as follows. 
In the next section, the time-dependent variational approach to the pure 
Yang-Mills theory, especially the $su(3)$ gauge theory as the QCD, is 
formulated in the Hamiltonian formalism. 
In \S 3, the time-dependent variational equations for the quantum 
gluon fields are reformulated in a form of the Liouville-von Neumann 
equation for the reduced density matrix of the quantum fluctuation fields 
at zero and the finite temperatures. 
In \S 4, the shear viscosity is evaluated in our framework for 
pure gluonic matter by using the linear response theory 
from the viewpoint of weakly coupled QCD or weakly correlated pure 
gluonic matter. 
The last section is devoted to a summary and concluding remarks.

\section{Time-dependent variational approach to QCD}

In this section, we give a variational method for the 
pure Yang-Mills gauge theory with color $su(3)$ symmetry 
in the functional Schr\"odinger picture 
with a Gaussian approximation, which 
is developed in Ref.\citen{Dominique}, in a slightly different 
manner. 
The trial state is constructed, paying an attention to the 
canonicity condition\cite{MMSK,YK} in our time-dependent variational 
approach. 
As a result, the equations of motion for variational functions 
are obtained as canonical equations of motion in classical mechanics.

\subsection{Hamiltonian formalism of pure gauge theory}

In this subsection, we summarize the Hamiltonian formalism 
of the pure Yang-Mills gauge theory for the sake of the 
definiteness of notations.

Let us start with the following Lagrangian density for the 
pure gauge theory with the color $su(N)$ symmetry: 
\beq\label{2-1}
& &{\cal L}=-\frac{1}{4}F_{\mu\nu}^aF_a^{\mu\nu}\ , \nonumber\\
& & \ F_{\mu\nu}^a=\partial_{\mu}A_{\nu}^a-\partial_\nu A_{\mu}^a
+gf_{abc}A_{\mu}^b A_{\nu}^c \ , 
\eeq
where $A_{\mu}^a$ represents the gauge field and the Greek indices such as 
$\mu$, $\nu$ etc. and 
the Roman indices such as $a$, $b$ etc. 
mean the Lorentz and the color indices, respectively. 
The repeated indices are summed up. 
Later, we use another Roman indices such as $i$, $j$, $\cdots$, which 
means the space components of the Lorentz indices, that is 1, 2 and 3. 
Here, $g$ represents the coupling constant and $f_{abc}$ is 
the structure constant for the color $su(N)$: 
\b\label{2-2}
[ \ T_a \ , \ T_b\ ]=if_{abc}T_c \ , 
\end{equation}
where $\{ T_a \}$ is the $su(N)$ generators. 
In the adjoint representation, the $su(N)$ generator can be expressed as 
$(T_a)_b^c=-if_{abc}$

The conjugate momentum, $\pi^{a\mu}$, for the field $A_\mu^a$ is defined as 
\b\label{2-3}
\pi^{a\mu}=\frac{\partial {\cal L}}{\partial {\dot A}_\mu^a}
=F^{a\mu} \ .
\end{equation}
Here, ${\dot A}=\partial A/\partial t$. 
We introduce the vector notation such as 
${\mib A}_a=(A_a^1, A_a^2, A_a^3)$. Then, the conjugate momentum 
with a space component can be expressed as 
\beq\label{2-4}
{\mib \pi}^a&=&-(F_{10}^a, F_{20}^a, F_{30}^a) \nonumber\\
&=&-{\dot {\mib A}}^a-\nabla A_0^a+gf_{abc}{\mib A}^bA_0^c \nonumber\\
&=&{\mib E}^a \ .
\eeq
Here, we define the color electric field ${\mib E}^a$. 
As is similar to the color electric fields, we define the color 
magnetic field as 
\beq\label{2-5}
{\mib B}^a&=&-(F_{23}^a, F_{31}^a, F_{12}^a) \nonumber\\
&=&\nabla \times {\mib A}^a-\frac{1}{2}gf_{abc}{\mib A}^b \times 
{\mib A}^c \ .
\eeq
Thus, we define the Hamiltonian density ${\cal H}_0$ as 
\beq\label{2-6}
{\cal H}_0&=&\pi^{a\mu}{\dot A}_\mu^a-{\cal L} \nonumber\\
&=&\frac{1}{2}\left[({\mib E}^a)^2+({\mib B}^a)^2\right]
+{\mib \pi}^a\cdot (\nabla A_0^a
-gf_{abc}{\mib A}^b A_0^c) \ .
\eeq
As is well known, the gauge theory leads to the constrained system. 
Namely, the conjugate momentum $\pi^a_0$ is identical to zero, so 
it is necessary to impose a constraint condition 
and the consistency condition for the time evolution as 
\beq\label{2-7}
& &\pi_0^a=F^{a,00}=0 \ , \qquad 
{\dot \pi}_0^a=0 \ .
\eeq
This fact leads to the Dirac theory of constrained system. 
In terms of the analytic mechanics, the constrained condition is written 
as
\beq\label{2-8}
& &{\dot \pi}_0^a=\{ \ \pi_0^a \ , \ \int d^3{\mib x}{\cal H} \ \}_P
={\mib D}\cdot {\pi}^a =0 \ , \nonumber\\
& &{\mib D}\cdot {\mib \pi}^a=\nabla\cdot {\mib \pi}^a
-gf_{abc}{\mib A}^b\cdot {\mib \pi}^c \nonumber\\ 
& &\qquad\quad
=\nabla\cdot {\mib E}^a
+ig\cdot if_{abc}{\mib A}^b\cdot {\mib E}^c \ , 
\eeq
where $\{ \ ,\ \}_P$ represents the Poisson bracket. 
Thus, the Hamiltonian is written as 
\beq\label{2-9}
\int d^3{\mib x}{\cal H}_0&=&\int d^3{\mib x}
\frac{1}{2}\left[({\mib E}^a)^2+({\mib B}^a)^2\right]
+\int d^3{\mib x}({\mib \pi}^a\cdot \nabla A_0^a-
gf_{abc}{\mib \pi}^a\cdot{\mib A}^b A_0^c) \nonumber\\
&=&\int d^3{\mib x}
\frac{1}{2}\left[({\mib E}^a)^2+({\mib B}^a)^2\right] \ ,
\eeq
where we used the integrated by part in the second term and 
the constrained condition (\ref{2-7}). 
Thus, hereafter, we use the Hamiltonian density as 
\b\label{2-10}
{\cal H}_0=\frac{1}{2}\left[({\mib E}^a)^2+({\mib B}^a)^2\right] \ .
\end{equation}

For the later convenience, 
we introduce the following variable, ${\cal G}$: 
\beq\label{2-11}
{\cal G}&=&{\cal G}^a T^a\nonumber\\
&=&\left(\nabla\cdot {\mib E}^a
+ig\cdot if_{abc}{\mib A}^b\cdot {\mib E}^c\right)T^a \nonumber\\
&=&\nabla\cdot{\mib E}+ig[\ A^i\ , \ E^i \ ] ,
\eeq
where $A^i=A_a^i T^a$ and so on, and 
${\cal G}^a={\mib D}\cdot {\mib \pi}^a$. 
Thus, it is understood that 
${\cal G}$ is nothing but the infinitesimal generator of the 
gauge transformation.

\subsection{Variational approach to pure gauge theory in quantum 
field theory}

In this subsection, we formulate the time-dependent variational method 
for the pure Yang-Mills gauge theory by using the functional 
Schr\"odinger picture\cite{JK,KV} within the Gaussian approximation. 
We formulate our variational method in the canonical form 
by the help of the canonical variable or canonicity conditions.\cite{MMSK,YK}

The time-dependent variational principle is formulated as 
\b\label{2-12}
\delta\int dt \bra{\Phi}i\frac{\partial}{\partial t}-\int d^3{\mib x}
{\cal H} \ket{\Phi}=0 \ ,
\end{equation}
where ${\cal H}$ means the Hamiltonian density under consideration. 
In the functional Schr\"odinger picture, the commutation relation 
$[\ A_i^a({\mib x}) \ , \ E_j^b({\mib y}) \ ]=i\delta_{ij}\delta_{ab}
\delta^3({\mib x}-{\mib y})$ leads to 
\b\label{2-13}
E_i^a({\mib x})\ket{\Phi}=-i\frac{\delta}{\delta A_i^a({\mib x})}\ket{\Phi}
\ .
\end{equation}

It is restricted ourselves that the trial state $\ket{\Phi}$ or the 
trial wave functional $\Phi({\mib A}^a)=\langle {\mib A}^a\ket{\Phi}$ has the 
following 
Gaussian form as 
\b\label{2-14}
\Phi({\mib A}^a)={\cal N}^{-1} \exp(i\langle {\ovl {\mib E}}\ket{{\mib A}-
{\ovl {\mib A}}})\exp\left(
-\bra{{\mib A}-{\ovl {\mib A}}}
\frac{1}{4G}-i\Sigma\ket{{\mib A}-{\ovl {\mib A}}}
\right) \ .
\end{equation}
Here, we used abbreviated notations such as 
\beq\label{2-15}
& &\langle {\ovl {\mib E}}\ket{{\mib A}}=\int d^3{\mib x}
{\ovl {\mib E}}^a({\mib x},t)\cdot{\mib A}^a({\mib x}) \ , \nonumber\\
& &\bra{{\mib A}}\frac{1}{4G}\ket{{\mib A}}
=\int\int d^3{\mib x} d^3{\mib y}A^a_i({\mib x}) \frac{1}{4}G^{-1}{}^{ab}_{ij}
({\mib x},{\mib y},t)A^b_j({\mib y}) \ .
\eeq
Here, ${\ovl A}_i^a({\mib x},t)$, ${\ovl E}_i^a({\mib x},t)$, 
$G_{ij}^{ab}({\mib x},{\mib y},t)$ and 
$\Sigma_{ij}^{ab}({\mib x},{\mib y},t)$ are the variational functions 
which are determined by the time-dependent variational principle. 
The reason why the form (\ref{2-14}) is adopted is that the 
canonicity conditions for $({\ovl A}_i^a, {\ovl E}_i^a)$ and 
$(G_{ij}^{ab}, \Sigma_{ij}^{ab})$ are automatically satisfied: 
\beq\label{2-16}
& &\bra{\Phi}i\frac{\delta}{\delta {\ovl A}_i^a}\ket{\Phi}={\ovl E}_i^a \ ,
\qquad
\bra{\Phi}i\frac{\delta}{\delta {\ovl E}_i^a}\ket{\Phi}=0 \ , \nonumber\\
& &\bra{\Phi}i\frac{\delta}{\delta G_{ij}^{ab}}\ket{\Phi}=0 \ ,
\qquad
\bra{\Phi}i\frac{\delta}{\delta \Sigma_{ij}^{ab}}\ket{\Phi}=-G_{ij}^{ab} \ . 
\eeq
Thus, our time-dependent variational method is formulated as 
a canonical form.

In the functional Schr\"odinger picture, the expectation values are 
easily calculated such as follows: 
\beq\label{2-17}
& &\bra{\Phi}A_i^a({\mib x})\ket{\Phi}={\ovl A}_i^a({\mib x},t) \ , 
\nonumber\\
& &\bra{\Phi}E_i^a({\mib x})\ket{\Phi}={\ovl E}_i^a({\mib x},t) \ , 
\nonumber\\
& &\bra{\Phi}A_i^a({\mib x})A_j^b({\mib y})\ket{\Phi}=
{\ovl A}_i^a({\mib x},t){\ovl A}_j^b({\mib y},t)
+G_{ij}^{ab}({\mib x},{\mib y},t) \ , 
\nonumber\\
& &\bra{\Phi}E_i^a({\mib x})E_j^b({\mib y})\ket{\Phi}=
{\ovl E}_i^a({\mib x},t){\ovl E}_j^b({\mib y},t)
+\frac{1}{4}G^{-1}{}_{ij}^{ab}({\mib x},{\mib y},t)
+4(\Sigma G \Sigma)_{ij}^{ab}({\mib x},{\mib y},t) \ , \nonumber\\
& &\bra{\Phi}{\mib A}^a({\mib x})\cdot {\mib E}^b({\mib x})\ket{\Phi}=
{\ovl {\mib A}}^a({\mib x},t)\cdot {\ovl {\mib E}}^b({\mib x},t)
+2(G \Sigma)_{ii}^{ab}({\mib x},{\mib x},t) \ . 
\eeq
Thus, it is understood that 
${\ovl {A}}_i^a$ represent the classical fields of gauge fields and 
the diagonal component of $G_{ij}^{ab}$, that is, $G_{ii}^{aa}$, 
where indices $i$ and $a$ are no sum,  
is a quantum fluctuations around the classical field ${\ovl {A}}_i^a$. 
Thus, in this functional Schr\"odinger picture, the two-point 
function $G_{ij}^{ab}({\mib x},{\mib y},t)$ plays a role of 
the gauge-particle propagator.

It should be noted here that 
the trial state (\ref{2-14}) does not have the gauge symmetry, 
that is ${\cal G}\ket{\Phi}\neq 0$. 
Thus, we impose the gauge invariance by introducing the Lagrange multiplier. 
From (\ref{2-8}), the constraint ${\mib D}\cdot{\mib \pi}^a=0$ 
is recast into another form ${\cal G}^a=0$ from (\ref{2-11}), 
where ${\cal G}^a$ is the generator of the gauge transformation. 
%From (\ref{2-11}), the generator of the gauge transformation ${\cal G}$ 
%is identical to the constraint ${\mib D}\cdot {\mib \pi}^a T^a$. 
Thus, we introduce the effective Hamiltonian density ${\cal H}$ by 
considering the gauge invariance in the space of the trial states 
as\cite{Dominique} 
\b\label{2-18}
{\cal H}={\cal H}_0-\omega^a({\mib x}){\cal G}^a({\mib x}) \ , 
\end{equation}
where $\omega^a({\mib x})$ represents a Lagrange multiplier, which 
insure the constraint ${\cal G}^a({\mib x})=0$. 
Thus, we use the above Hamiltonian density in order to determine the 
time dependences of the variational functions 
${\ovl A}_i^a({\mib x},t)$, ${\ovl E}_i^a({\mib x},t)$, 
$G_{ij}^{ab}({\mib x},{\mib y},t)$ and 
$\Sigma_{ij}^{ab}({\mib x},{\mib y},t)$.

The expectation value of the Hamiltonian can be expressed as 
following simple form: 
\beq
\langle H \rangle&=&
\bra{\Phi}\int d^3{\mib x}\left[
{\cal H}_0-\omega^a({\mib x}){\cal G}^a({\mib x})
\right]
\ket{\Phi} \nonumber\\
&=&\langle H_0 \rangle-\int d^3{\mib x}\omega^a({\mib x})
\langle {\cal G}^a({\mib x}) \rangle \ , \nonumber\\
\langle H_0 \rangle&=&
\int d^3{\mib x}\biggl(
\frac{1}{2}{\ovl {\mib B}}^a({\mib x})\cdot {\ovl {\mib B}}^a({\mib x})
+\frac{1}{2}{\ovl {\mib E}}^a({\mib x})\cdot {\ovl {\mib E}}^a({\mib x})
+\frac{1}{8}{\rm Tr}\bra{\mib x}G^{-1}\ket{\mib x} \nonumber\\
& &\qquad\quad
+2{\rm Tr}\bra{\mib x}\Sigma G \Sigma \ket{\mib x}+
\frac{1}{2}{\rm Tr}\bra{\mib x}KG\ket{\mib x}
+\frac{g^2}{8}
\left({\rm Tr}[S^i T^a\bra{\mib x}G\ket{\mib x}]\right)^2 \nonumber\\
& &\qquad\quad
+\frac{g^2}{4}{\rm Tr}\left[S^iT^a\bra{\mib x}G\ket{\mib x}
S^iT^a\bra{\mib x}G\ket{\mib x}\right]\biggl) \ , \nonumber\\
\langle {\cal G}^a({\mib x})\rangle&=&
\nabla\cdot {\ovl {\mib E}}^a({\mib x})-ig{\rm Tr}
\bra{\mib x}T^a[\Sigma\ , \ G]\ket{\mib x}
+ig\cdot if_{abc}{\ovl {\mib A}}^b\cdot {\ovl {\mib E}}^c \ , 
\label{2-19}
\eeq
where we define 
\beq\label{2-20}
& &{\ovl B}_i^a=\epsilon_{ijk}\partial_j {\ovl A}_k^a
-\frac{1}{2}gf^{abc}\epsilon_{ijk}{\ovl A}_j^b{\ovl A}_k^c \ , \nonumber\\
& &(S^i)_{jk}=i\epsilon_{ijk}\ , \qquad 
(T^a)^{bc}=-if^{abc} \ , \nonumber\\
& &K=(-i{\mib S}\cdot {\mib D})^2-g{\mib S}\cdot{\ovl {\mib B}} \ , \nonumber\\
& &{\mib D}=\nabla-ig{\ovl {\mib A}} \ , \qquad
{\ovl {\mib A}}={\ovl A}_i^a T^a \ , \quad 
{\ovl {\mib B}}={\ovl {\mib B}}^a T^a \ .
\eeq
Here, we can use the abbreviated notation such as
$\bra{\mib x}G\ket{\mib y}=G_{ij}^{ab}({\mib x},{\mib y},t)$. 
In the above representation, ${\mib S}$ represent the spin 1 matrices 
whose spatial component with $i$ is $S^i$.

\subsection{Variational equations and their solutions in the 
time-independent case}

The equations of motion for the variational functions are derived 
from the time-dependent variational principle in Eq.(\ref{2-12}) with 
the Hamiltonian density (\ref{2-18}). The results are summarized 
in the form of canonical equations of motion as 
\bsub\label{2-21}
\beq
& &{\dot {\ovl {\mib A}}}^a({\mib x},t)
=\frac{\delta \langle H \rangle}{\delta {\ovl {\mib E}}^a({\mib x},t)}\ , 
\qquad
{\dot {\ovl {\mib E}}}^a({\mib x},t)
=-\frac{\delta \langle H \rangle}{\delta {\ovl {\mib A}}^a({\mib x},t)} \ , 
\label{2-21a}\\
& &{\dot G}_{ij}^{ab}({\mib x},{\mib y},t)
=\frac{\delta \langle H \rangle}{\delta \Sigma_{ij}^{ab}({\mib x},{\mib y},t)} 
\ , \qquad
{\dot \Sigma}_{ij}^{ab}({\mib x},{\mib y},t)
=-\frac{\delta \langle H \rangle}{\delta G_{ij}^{ab}({\mib x},{\mib y},t)} 
\ , 
\label{2-21b}
\eeq
\esub
where $\langle H \rangle$ is given in Eq.(\ref{2-19}). 

In the time-independent case, the above equations of motion in 
Eq.(\ref{2-21a}) 
allow the following solutions within the lowest order 
of $g$ as 
\b\label{2-22}
{\ovl {\mib A}}^a({\mib x})=0 \ , \qquad 
{\ovl {\mib E}}^a({\mib x})=-\nabla \omega^a({\mib x}) \ .
\end{equation}
In the above solutions, the mean field or the classical field 
${\overline {\mib A}}^a({\mib x})$ is identical to zero,
namely, under this situation, only the quantum field 
is dealt with in this matter system. 
Further, in the lowest order of $g$, 
the time-independent equations (\ref{2-21b}) present the following solutions 
under ${\ovl {\mib A}}^a=0$ : 
\bsub\label{2-23}
\beq
G_{ij}^{ab}({\mib x},{\mib y})
&=&\int\frac{d^3{\mib k}}{(2\pi)^3}e^{i{\mib k}\cdot({\mib x}-{\mib y})}
G_{ij}^{ab}({\mib k})
=\delta^{ab}
\int\frac{d^3{\mib k}}{(2\pi)^3}e^{i{\mib k}\cdot({\mib x}-{\mib y})}
G_k \left(\delta_{ij}-\frac{k_ik_j}{{\mib k}^2}
\right) \ , \nonumber\\
& &G_{ij}^{ab}({\mib k})=\delta^{ab}\left(\delta_{ij}-\frac{k_ik_j}{{\mib k}^2}
\right)G_k\ , \qquad
G_k=\frac{1}{2|{\mib k}|} \ , 
\label{2-23a}\\
\Sigma_{ij}^{ab}({\mib x},{\mib y})&=&
\int\int \frac{d^3{\mib k}}{(2\pi)^3}\frac{d^3{\mib k}'}{(2\pi)^3}
e^{i{\mib k}'\cdot{\mib x}}
e^{-i{\mib k}\cdot{\mib y}}\bra{{\mib k}'a}\Sigma_{ij}
\ket{b{\mib k}} \ , \nonumber\\
& &\bra{{\mib k}'a}\Sigma_{ij}\ket{b{\mib k}}
=\frac{1}{2}\left(\delta_{il}-\frac{k'_ik'_l}{{\mib k'}^2}\right)
\left(\delta_{lj}-\frac{k_lk_j}{{\mib k}^2}\right)
g\omega^c({\mib q})\delta^3({\mib k'}-{\mib k}-{\mib q}) \nonumber\\
& &\qquad\qquad\qquad\quad
\times f^{abc}
\left(\frac{G_k-G_{k'}}{G_k+G_{k'}}\right)\ , 
\label{2-23b}\\
& &\qquad
\omega^a({\mib q})=\int d^3{\mib x} \omega^a({\mib x})
e^{-i{\mib k}\cdot{\mib x}} \ . 
\eeq
\esub
Thus, $G_{ij}^{ab}({\mib k})$ has only the transverse component. 
This feature is plausible for the gauge-particle propagation.

\subsection{Thouless-Valatin correction}

In the Hamiltonian density (\ref{2-18}), the constrained term 
$\omega^a{\cal G}^a$ is introduced. 
This treatment is resemble to that of the nuclear rotation.\cite{RS} 
In the nuclear many-body theory, the collective rotational motion 
of the axially symmetric deformed nuclei for $z$-axis 
is described in the same way used in 
this section. 
If the nuclear rotation occurs in the perpendicular to the $x$ axis, 
the state $\ket{\Psi(t)}=e^{-iEt}\ket{\Psi}$ is replaced as 
\b\label{2-24}
\ket{\Psi_\omega(t)}=e^{-i\omega t{\hat J}_x}e^{-iE_{\omega}t}\ket{\Psi} \ , 
\end{equation}
where the angular momentum operator ${\hat J}_x$ and the angular velocity 
$\omega$ are introduced. 
Then, the Schr\"odinger equation, $i\partial_t\ket{\Psi_\omega(t)}
={\hat H}\ket{\Phi_\omega(t)}$ is recast into 
\b\label{2-25}
({\hat H}-\omega{\hat J}_x)\ket{\Psi}=E_\omega\ket{\Psi} \ ,
\end{equation}
where ${\hat H}$ is the original nuclear Hamiltonian. 
It is known that we have to get rid of the effect of nuclear rotation 
from the total energy as 
\b\label{2-26}
E=\bra{\Psi}{\hat H}\ket{\Psi}-\frac{\bra{\Psi}{\hat J}_x^2\ket{\Psi}}
{2I} 
=\bra{\Psi}{\hat H}\ket{\Psi}-\Delta E_{\rm TV}\ , 
\end{equation}
where $I$ is the moment of inertia and is defined as\cite{I} 
\b\label{2-27}
I=\lim_{\omega \rightarrow 0}\frac{\bra{\Psi_\omega}{\hat J}_x
\ket{\Psi_\omega}}
{\omega} \ .
\end{equation}
This energy correction, $\Delta E_{\rm TV}$, in Eq.(\ref{2-26}) 
is well known as the Thouless-Valatin correction.\cite{TV} 

Thus, in the approach to the pure Yang-Mills theory, 
it is first pointed out that 
the same correction term is necessary in Ref.\citen{Dominique}. 
For the pure Yang-Mills theory in the treatment of the variational 
method, the Thouless-Valatin correction term can be expressed as 
\b\label{2-28}
\Delta E_{\rm TV}=\int\int d^3{\mib x}d^3{\mib y}
\bra{\Phi}{\cal G}^a({\mib x}){\cal G}^b({\mib y})\ket{\Phi}
\bra{a{\mib x}}\frac{1}{2{\cal I}}\ket{b{\mib y}} \ ,
\end{equation}
where the moment of inertia for the gauge rotation, 
${\cal I}^{ab}({\mib x},{\mib y})=\bra{a{\mib x}}{\cal I}\ket{b{\mib y}}$, 
is defined as 
\b\label{2-29}
{\cal I}^{ab}({\mib x},{\mib y})=\lim_{\omega^b({\mib y})\rightarrow 0}
\frac{\bra{\Phi}{\cal G}^a({\mib x})\ket{\Phi}}{\omega^b({\mib y})} \ .
\end{equation}
It is first shown that, in Ref.\citen{Dominique}, 
the above Thouless-Valatin correction term and the contribution of the 
moment of inertia for the gauge rotation play essential roles 
in order to reproduce the 
one-loop running coupling constant. 
The validity of our time-dependent variational approach owes the fact 
that the one-loop running coupling constant 
is exactly reproduced in the lowest order 
approximation of $g$ under ${\ovl {\mib A}}^a=0$ developed in 
Ref.\citen{Dominique}.

\section{Time-dependent variational equations for quantum gauge fields}

In this section, we present the equations of motion for the quantum 
fluctuations around the classical field configurations ${\ovl {\mib A}}^a$ 
and ${\ovl {\mib E}}^a$, namely, $G_{ij}^{ab}$ and $\Sigma_{ij}^{ab}$ 
for quantum gauge fields, 
in a slightly different forms from Eq.(\ref{2-21b}). 
We can formulate the equations of motion for quantum gauge fields 
as the Liouville-von Neumann equation.

\subsection{Liouville-von Neumann equation for quantum gauge fields}

First, the reduced density matrix ${\cal M}$ is introduced 
as is similar to the Hartree-Bogoliubov theory for many-body physics 
in the boson systems. 
We define the reduced density matrix\cite{TVM99,TVM00} 
for the quantum gauge fields as 
\beq\label{3-1}
{\cal M}_{ij}^{ab}({\mib x},{\mib y},t)&=&
\left(\begin{array}{@{\,}cc@{\,}}
-i\langle {\hat A}_i^a({\mib x},t){\hat E}_j^b({\mib y},t)\rangle
-\frac{1}{2} & 
\langle {\hat A}_i^a({\mib x},t){\hat A}_j^b({\mib y},t)\rangle \\
\langle {\hat E}_i^a({\mib x},t){\hat E}_j^b({\mib y},t)\rangle & 
i\langle {\hat E}_i^a({\mib x},t){\hat A}_j^b({\mib y},t)\rangle-\frac{1}{2}
\end{array}\right) \nonumber\\
&=&
\left(\begin{array}{@{\,}cc@{\,}}
-2i(G\Sigma)_{ij}^{ab}({\mib x},{\mib y},t) & 
G_{ij}^{ab}({\mib x},{\mib y},t) \\
\frac{1}{4}(G^{-1})_{ij}^{ab}({\mib x},{\mib y},t)
+4(\Sigma G\Sigma)_{ij}^{ab}({\mib x},{\mib y},t) & 
2i(\Sigma G)_{ij}^{ab}({\mib x},{\mib y},t)
\end{array}\right) \ , \nonumber\\
{\hat A}_i^a({\mib x},t)&=&A_i^{a}({\mib x})-\langle A_i^{a}({\mib x})
\rangle=A_i^a({\mib x})-{\ovl A}_i^a({\mib x},t) \ , \nonumber\\
{\hat E}_i^a({\mib x},t)&=&E_i^{a}({\mib x})-\langle E_i^{a}({\mib x})
\rangle=E_i^a({\mib x})-{\ovl E}_i^a({\mib x},t) \ , 
\eeq
where the symbol $\langle \cdots \rangle$ represents the expectation values 
for the state $\ket{\Phi(t)}$in (\ref{2-14}) as is shown in Eq.(\ref{2-17}). 
In the later, the expectation values are replaced into the thermal averages. 
Here, ${\hat A}_i^a$ means the quantum fluctuations around the classical 
configuration ${\ovl A}_i^a$. 
Thus, this reduced density matrix ${\cal M}$ can be regarded 
as the one consist by the quantum gauge fields. 

By the help of the Heisenberg equations of motion in the 
Heisenberg picture, 
the time evolution of the reduced density matrix is easily derived. 
As the result, we can obtain the 
following Liouville-von Neumann type equation of motion for the 
reduced density matrix composed of the quantum gauge fields as 
\bsub\label{3-2}
\beq
i{\dot {\cal M}}_{ij}^{ab}({\mib x},{\mib y},t)
&=&[\ {\wtilde {\cal H}} \ , \ {\cal M}\ ]_{ij}^{ab}({\mib x},{\mib y},t) \ ,
\label{3-2a}\\
{\wtilde {\cal H}}_{ij}^{ab}({\mib x},{\mib y},t)
&=&
\left(\begin{array}{@{\,}cc@{\,}}
\lambda_{ij}^{ab}({\mib x}) & 
\delta_{ij}\delta^{ab} \\
\Gamma_{ij}^{ab}({\mib x},t) & 
\lambda_{ij}^{ab}({\mib x})
\end{array}\right)\delta^3({\mib x}-{\mib y}) \ , 
\label{3-2b}\\
\Gamma_{ij}^{ab}({\mib x},t)&=&
K_{ij}^{ab}+g^2\left(
S_kT^c\bra{{\mib x}}G\ket{\mib x}S_kT^c\right)_{ij}^{ab} 
%\nonumber\\
%& &
+\frac{g^2}{2}\left(S_kT^c\right)_{ij}^{ab}{\rm Tr}\left[
S_kT^c\bra{\mib x}G\ket{\mib x}\right]
\nonumber\\
\lambda_{ij}^{ab}({\mib x})&=&
-ig\omega^p({\mib x})f^{pab}\delta_{ij}=g(\omega^p({\mib x}) T^p)_{ij}^{ab} \ ,
\label{3-2c}
\eeq
\esub
where $K_{ij}^{ab}$, $S_k$ and $T^c$ have been defined in Eq.(\ref{2-20}). 
Here, ${\wtilde {\cal H}}$ is the Hamiltonian matrix which governs 
the time evolution of the reduced density matrix. 

At this stage, it is important to indicate that the 
square of the reduced density matrix ${\cal M}$ 
is easily obtained from the second line of Eq.(\ref{3-1}) as
\b\label{3-3}
{\cal M}^2=\left(\begin{array}{@{\,}cc@{\,}}
\frac{1}{4} & 0 \\
0 & \frac{1}{4}
\end{array}\right) \ . 
\end{equation}
Thus, the eigenvalues of the reduced density matrix are $\pm 1/2$ 
as is similar to the case of the linear sigma model.\cite{TVM99,TVM00}

The eigenvector for the eigenvalue $1/2$ can be expressed as 
\b\label{3-4}
\bra{\mib x}1/2_n ai\rangle=\left(\begin{array}{@{\,}c@{\,}}
u_{n}{}_i^a({\mib x},t) \\
v_{n}{}_i^a({\mib x},t) 
\end{array}\right) \ , 
\end{equation}
where $n$ represents a certain quantum number. 
Then the following eigenvalue equation should be satisfied:
\b\label{3-5}
\int d^3{\mib y} {\cal M}_{ij}^{ab}({\mib x},{\mib y},t)
\left(\begin{array}{@{\,}c@{\,}}
u_{n}{}_j^b({\mib y},t) \\
v_{n}{}_j^b({\mib y},t) 
\end{array}\right) =\frac{1}{2}
\left(\begin{array}{@{\,}c@{\,}}
u_{n}{}_i^a({\mib y},t) \\
v_{n}{}_i^a({\mib y},t) 
\end{array}\right)  \ .
\end{equation}
From the above eigenvalue equation, we can derive the following equation:
\b\label{3-6}
\int d^3{\mib y} {\cal M}_{ij}^{ab}({\mib x},{\mib y},t)
\left(\begin{array}{@{\,}c@{\,}}
u_{n}^*{}_j^b({\mib y},t) \\
-v_{n}^*{}_j^b({\mib y},t) 
\end{array}\right) =-\frac{1}{2}
\left(\begin{array}{@{\,}c@{\,}}
u_{n}^*{}_i^a({\mib y},t) \\
-v_{n}^*{}_i^a({\mib y},t) 
\end{array}\right)  \ .
\end{equation}
Thus, we conclude that the ${}^t(u_n^*, -v_n^*)$ is the eigenvector 
for the reduced density matrix ${\cal M}$ with the eigenvalue $-1/2$, 
which is expressed as $\bra{\mib x}-\!1/2_nai\rangle$. 
Further, we can derive the following: 
\b\label{3-7}
\int d^3{\mib y} {\cal M}_{ij}^{ab}({\mib x},{\mib y},t)^{\dagger}
\left(\begin{array}{@{\,}c@{\,}}
v_{n}{}_j^b({\mib y},t) \\
u_{n}{}_j^b({\mib y},t) 
\end{array}\right) =\frac{1}{2}
\left(\begin{array}{@{\,}c@{\,}}
v_{n}{}_i^a({\mib y},t) \\
u_{n}{}_i^a({\mib y},t) 
\end{array}\right)  \ .
\end{equation}
Thus, we can introduce another vector as 
\b\label{3-8}
\bra{\mib x}1/2_n^{\dagger} ai\rangle=\left(\begin{array}{@{\,}c@{\,}}
v_{n}{}_i^a({\mib x},t) \\
u_{n}{}_i^a({\mib x},t) 
\end{array}\right) \ , 
\end{equation}
Then, we can express 
$\bra{1/2_n^{\dagger} ai}{\mib x}\rangle
=(v_n^*{}_i^a({\mib x},t) , u_n^*{}_i^a({\mib x},t))$.

\subsection{Spectral decomposition of reduced density matrix}

In the previous subsection, it is learned that the reduced density matrix 
${\cal M}$ has the eigenvalues $\pm 1/2$. 
In this subsection, the eigenvalue equation is summarized 
taking into account the extension to the finite temperature systems. 
Further, by using the eigenstates for ${\cal M}$, the reduced density 
matrix is expressed in the form of the spectral decomposition. 

Considering the extension to finite temperature systems, 
the eigenvalue equations are described with the abstract representation as 
\beq\label{3-9}
& &{\cal M}_{ij}^{ab}\ket{\sigma f_n,b,j}=\sigma f_n\ket{\sigma f_n,a,i} \ , 
\qquad
{\cal M}_{ij}^{ab}{}^{\dagger}\ket{\sigma f_n^{\dagger},b,j}
=\sigma f_n\ket{\sigma f_n^{\dagger},a,i} \ , 
\eeq
where $f=1/2$ and $\sigma=\pm$ 
at zero temperature developed in the previous subsection. 
From the second equation in (\ref{3-9}), we obtain 
\b\label{3-10}
\bra{\sigma f_n^{\dagger},b,j}{\cal M}_{ji}^{ba}
=\sigma f_n\bra{\sigma f_n^{\dagger},a,i} \ .
\end{equation}
By using Eqs.(\ref{3-9}) and (\ref{3-10}), we can easily derive the 
following orthogonal relations as 
\beq\label{3-11}
& &\sum_{a,i}\bra{\sigma' f_{n'}^{\dagger},a,i} \sigma f_n,a,i\rangle
=\delta_{\sigma'\sigma}\delta_{n'n} \ , 
\eeq
where the normalization 
condition of $\ket{\sigma f_n,a,i}$ is taken into account. 
By using the completeness relation $\int d^3{\mib x}
\ket{\mib x}\bra{\mib x}=1$ and the expression such as (\ref{3-4}) with 
$\sigma=\sigma'=1$ or $\sigma=1$ and $\sigma'=-1$, 
the above condition (\ref{3-11}) is rewritten as\cite{TVM99} 
\beq\label{3-12}
& &\sum_{a,i}\int d^3{\mib x}\left(v^*_{n'}{}_i^a({\mib x})
u_{n}{}_i^a({\mib x})+u^*_{n'}{}_i^a({\mib x})v_{n}{}_i^a({\mib x})
\right)=\delta_{nn'} \ , 
\nonumber\\
& &\sum_{a,i}\int d^3{\mib x}\left(u^*_{n'}{}_i^a({\mib x})
v^*_{n}{}_i^a({\mib x})-v^*_{n'}{}_i^a({\mib x})u^*_{n}{}_i^a({\mib x})
\right)=0 \ , 
\nonumber\\
& &\sum_{a,i}\int d^3{\mib x}\left(u_{n'}{}_i^a({\mib x})
v_{n}{}_i^a({\mib x})-v_{n'}{}_i^a({\mib x})u_{n}{}_i^a({\mib x})
\right)=0 \ . 
\eeq
Thus, the reduced density 
matrix ${\cal M}$ itself can be expressed in terms 
of the eigenstates of ${\cal M}$ as follows:
\beq\label{3-13}
{\cal M}_{ij}^{ab}({\mib x},{\mib y},t)
&=&
\sum_{n (\sigma>0)} f_n
\biggl[ \left(\begin{array}{@{\,}c@{\,}}
u_{n}{}_i^a({\mib y},t) \\
v_{n}{}_i^a({\mib y},t) 
\end{array}\right)
(\ v_n^*{}_j^b({\mib y}) \ , \ u_n^*{}_j^b({\mib y})\ ) \nonumber\\
& &\qquad\qquad
+\left(\begin{array}{@{\,}c@{\,}}
u^*_{n}{}_i^a({\mib y},t) \\
-v^*_{n}{}_i^a({\mib y},t) 
\end{array}\right)
(\ -v_n{}_j^b({\mib y}) \ , \ u_n{}_j^b({\mib y})\ )
\biggl] \ .
\eeq
We can easily verify that the above ${\cal M}$ satisfy 
the eigenvalue equations (\ref{3-5}) and (\ref{3-6}) with $f_n$ instead of 
$1/2$.

Next, let us determine the eigenvalue $f_n$ in the system at finite 
temperature. 
In the time-independent case, from Eq.(\ref{3-2a}), it is seen that 
the reduced density 
matrix and the Hamiltonian matrix commute each other. 
Thus, there exist the simultaneous eigenstates for ${\cal M}$ and 
${\cal {\wtilde H}}$ whose eigenvalues are $\sigma f_n$ and $\sigma E_n$, 
respectively. 
Namely, 
\beq\label{3-14}
& &\int d^3{\mib y} {\cal {\wtilde H}}_{ij}^{ab}({\mib x},{\mib y},t)
\left(\begin{array}{@{\,}c@{\,}}
u_{n}{}_j^b({\mib y},t) \\
v_{n}{}_j^b({\mib y},t) 
\end{array}\right) =E_n
\left(\begin{array}{@{\,}c@{\,}}
u_{n}{}_i^a({\mib y},t) \\
v_{n}{}_i^a({\mib y},t) 
\end{array}\right)  \ , \nonumber\\
& &\int d^3{\mib y} {\cal {\wtilde H}}_{ij}^{ab}({\mib x},{\mib y},t)
\left(\begin{array}{@{\,}c@{\,}}
u_{n}^*{}_j^b({\mib y},t) \\
-v_{n}^*{}_j^b({\mib y},t) 
\end{array}\right) =-E_n
\left(\begin{array}{@{\,}c@{\,}}
u_{n}^*{}_i^a({\mib y},t) \\
-v_{n}^*{}_i^a({\mib y},t) 
\end{array}\right)  \ .
\eeq
Thus, we can derive ${\rm Tr}({\cal {\wtilde H}}{\cal M})=\sum_m E_m$. 
Here, the Helmholtz free energy is defined as 
\beq\label{3-15}
F&=&\langle {\cal {\wtilde H}} \rangle -TS \ , \nonumber\\
& &\langle {\cal {\wtilde H}} \rangle=
{\rm Tr}({\cal {\wtilde H}}{\cal M})=\sum_m E_m \ , \nonumber\\
& &S=\sum_{m\sigma}\left[
(1+n_{m\sigma})\ln(1+n_{m\sigma})-n_{m\sigma}\ln n_{m\sigma}\right)\ , 
\eeq
where $T$ is temperature. Thus, the minimization condition is imposed:
\beq\label{3-16}
& &\delta F={\rm Tr}({\cal {\wtilde H}}\delta{\cal M})-T\delta S=0 \ .
\eeq
Here, we assume that $\sigma f_m$ depends on $n_{m\sigma}$ linearly. 
Under this assumption,\break 
$\delta {\cal M}/\delta n_{m\sigma}={}^t(u, v)(v^*, u^*)$ for $\sigma=+$ 
and $\delta {\cal M}/\delta n_{m\sigma}={}^t(u^*, -v^*)(-v, u)$ 
for $\sigma=-$ are obtained respectively. 
Thus, we obtain 
\beq\label{3-17}
& &\frac{\delta F}{\delta n_{m\sigma}}
=E_m-T\ln\frac{1+n_{m\sigma}}{n_{m\sigma}}=0 \ , \nonumber\\
& &\qquad {i.e.,}\qquad
n_{m\sigma}=\frac{1}{e^{E_m/T}-1} \ , 
\eeq
where $\sigma=\pm$. Thus, we omit the suffix $\sigma$ in $n_{m\sigma}$. 
Of course, the eigenvalue $f$ is reduced to $1/2$ when $T\rightarrow 0$. 
Thus, finally, we obtain 
\b\label{3-18}
f_m=n_m+\frac{1}{2} \ , \qquad 
n_m=\frac{1}{e^{E_m/T}-1} \ .
\end{equation}

From the (1,2)-component of the reduced density matrix in Eq.(\ref{3-1}), 
the two-point function $G$ can be decomposed by each spectral function as 
\b\label{3-19}
G_{ij}^{ab}({\mib x},{\mib y},t)=\sum_{n}f_n
\left[u_n{}_i^a({\mib x})u^*_n{}_j^b({\mib y})+
u_n^*{}_i^a({\mib x})u_n{}_j^b({\mib y})\right] \ .
\end{equation}
Hereafter, we take the quantum number $n$ as momentum ${\mib k}$. 
Then, we obtain 
\b\label{3-20}
G_{ij}^{ab}({\mib x},{\mib y},t)=\int \frac{d^3{\mib k}}{(2\pi)^3}
f_{\mib k}
\left[u_{\mib k}{}_i^a({\mib x})u^*_{\mib k}{}_j^b({\mib y})+
u_{\mib k}^*{}_i^a({\mib x})u_{\mib k}{}_j^b({\mib y})\right] \ .
\end{equation}
At zero temperature, we have already derived the expression of $G$ in 
Eq.(\ref{2-23}). In the above expression in (\ref{3-20}), 
$f_{\mib k}=1/2$ and $uu^*$ is also expressed. 
By using the knowledge of the zero temperature case, 
the above expression in (\ref{3-20}) can be recast into 
\beq\label{3-21}
G_{ij}^{ab}({\mib x},{\mib y},t)&=&\delta^{ab}
\int \frac{d^3{\mib k}}{(2\pi)^3}
G_k^T
\left(\delta_{ij}-\frac{k_ik_j}{{\mib k}^2}
\right)e^{i{\mib k}\cdot({\mib x}-{\mib y})} \ , \nonumber\\
G_k^T&=&(2n_{\mib k}+1)\frac{1}{2|{\mib k}|}\ .
\eeq

From the eigenvalue equation (\ref{3-14}), the energy eigenvalue 
for the quantum gauge fields is easily obtained. 
For simplicity, we neglect the gauge rotating term, namely, we 
put $\omega^a({\mib x})=0$. 
Then from (\ref{3-14}), we obtain 
\beq\label{3-22}
& &E_{\mib k}\delta_{ij}\delta^{ab}
=\Gamma^{1/2}{}_{ij}^{ab}({\mib k}) \ , \nonumber\\
& &\Gamma^{1/2}{}_{ij}^{ab}({\mib k})
=\delta^{ab}|{\mib k}|\left(\delta_{ij}-\frac{k_ik_j}{{\mib k}^2}\right) 
\eeq
up to the order of $g$.

Finally, it should be noted that 
the gauge-particle has only 
transverse component as is realized in Eq.(\ref{2-23}) 
with factor $(\delta_{ij}-k_ik_j/{\mib k}^2)$ . 
Since the reduced density matrix ${\cal M}$ includes $G$ and $\Sigma$ 
composed by the quantum gauge field, ${\cal M}$ also contains the 
factor $(\delta_{ij}-k_ik_j/{\mib k}^2)$. 
Thus, we define the following projection operator ${\hat P}$: 
\beq\label{3-23}
& &
\frac{\bra{{\mib k}'}{\hat P}_{ij}\ket{\mib k}}{\bra{\mib k'}{\mib k}\rangle}
=\delta_{ij}-\frac{k_ik_j}{{\mib k}^2}
\ .
\eeq
Then, ${\hat {\cal P}}$ certainly has a property of the projection operator, 
namely, ${\hat {\cal P}}^2={\hat {\cal P}}$. 
As is easily shown, the following relations are satisfied: 
\beq\label{3-24}
& &G={\hat {\cal P}}G{\hat {\cal P}}\ , \quad
\Sigma={\hat {\cal P}}\Sigma{\hat {\cal P}} \ , \nonumber\\
& &{\cal M}={\hat {\cal P}}{\cal M}{\hat {\cal P}}
={\hat {\cal P}}{\cal M}={\cal M}{\hat {\cal P}} \ .
\eeq
Since there always exists the projection factor to the transverse component, 
$(\delta_{ij}-k_ik_j/{\mib k}^2)$, 
the inverse of the two-point function $G$ should be regarded as the 
two point function which satisfies the following relation: 
\b\label{3-25}
G^{-1}G={\hat {\cal P}}\ . 
\end{equation}

\section{Transport coefficients of quantum gluonic matter}

Hereafter, we deal with the color $su(3)$ pure gauge theory, 
namely the QCD without quarks. 
In this section, we present the expression of the transport coefficients 
in our variational approach by using the Kubo formula\cite{Kubo,Mori} 
for the quantum gluonic matter with the color $su(3)$ symmetry. 

\subsection{Kubo formula based on the variational approach}

By taking into account 
an external source field ${\hat A}$ and its conjugate 
force $F(t)$, 
the Hamiltonian ${\hat H}$ is modified from ${\hat H}_0$ to 
\b\label{4-1}
{\hat H}(t)={\hat H}_0-{\hat A}F(t) \ . 
\end{equation}
If the external force $F(t)$ is adopted as $F(t)=F e^{-i\omega t}$, then, 
an observable ${\hat B}$ at time $t$,$\langle {\hat B} \rangle_t$, 
can be expressed in the linear response theory as 
\beq\label{4-2}
\langle {\hat B} \rangle_t &=&\langle {\hat B} \rangle_{\rm eq}
+\chi_{BA}(\omega) Fe^{-i\omega t} \ , 
\eeq
where $\langle \cdots \rangle_{\rm eq}$ means the thermal average 
with respect to the equilibrium state. 
Here, $\chi_{BA}(\omega)$ is called 
the complex admittance and is defined as 
\beq\label{4-3}
\chi_{BA}(\omega)&=&-\lim_{\epsilon\rightarrow +0}\frac{i}{\hbar}
\int_0^{\infty}dt \langle\ [\ {\hat A}\ , \ {\hat B}(t)\ ]\ \rangle_{\rm eq}
e^{i\omega t-\epsilon t}  \nonumber\\
&=&\lim_{\epsilon\rightarrow +0}\int_0^{\infty}dt
\varphi_{BA}(t)e^{i\omega t-\epsilon t} \ ,
\eeq
where 
\beq\label{4-4}
\varphi_{BA}(t)&=&-\frac{i}{\hbar}
\langle\ [\ {\hat A}\ , \ {\hat B}(t)\ ]\ \rangle_{\rm eq} \ \nonumber\\
&=&
\int_0^\beta d\lambda \langle {\dot {\hat A}}(-i\hbar\lambda)
{\hat B}(t)\rangle_{\rm eq}
\eeq
is a quantum response function. 
By using the integrated by part and an property of the equilibrium state, 
namely, $\langle {\dot A}B \rangle_{\rm eq}=
-\langle A{\dot B} \rangle_{\rm eq}$, the complex admittance is recast into 
\beq\label{4-5}
\chi_{BA}(\omega)
&=&\frac{1}{\hbar\omega}\lim_{\epsilon\rightarrow +0}
\int_0^{\infty}dt e^{i\omega t-\epsilon t}
\langle\ [\ {\hat B}(t)\ , \ {\dot {\hat A}}(0)\ ]\ \rangle_{\rm eq}
\nonumber\\
& &-\frac{1}{\hbar\omega}\lim_{\epsilon\rightarrow +0}
\int_0^{\infty}dt e^{-\epsilon t}
\langle\ [\ {\dot {\hat A}}(t)\ , \ {{\hat B}}(0)\ ]\ \rangle_{\rm eq}
\ . 
\eeq

Following the general theory,\cite{Mori} 
the transport coefficients are obtained by adopting both operators 
${\dot {\hat A}}$ and ${\hat B}$ being currents $J(t)$ as 
\beq\label{4-6}
\chi_{BA}(\omega,{\mib k}\!=\!{\mib 0})
&=&\frac{1}{\omega}\lim_{\epsilon\rightarrow +0}
\int_0^{\infty}dt \int d^3{\mib r} e^{i\omega t-\epsilon t}
\langle\ [\ J({\mib r},t)\ , \ J({\mib 0},0)\ ]\ \rangle_{\rm eq}
\nonumber\\
& &-\frac{1}{\omega}\lim_{\epsilon\rightarrow +0}
\int_0^{\infty}dt \int d^3{\mib r}e^{-\epsilon t}
\langle\ [\ J({\mib r},t)\ , \ J({\mib 0},0)\ ]\ \rangle_{\rm eq}
\ , 
\eeq
where we return to the natural unit, $\hbar=1$. 
Here, current $J(t)$ is defined as 
\beq\label{4-7}
& &J(t)=\int d^3{\mib r} J({\mib r},t)e^{-i{\mib k}\cdot {\mib r}} 
\eeq
for an application to the gluonic matter in mind.

\subsection{Shear viscosity in the quantum gluonic matter}

In the gluonic matter, the dependence of the coupling constant $g$ 
for the usual, not anomalous, 
shear viscosity is given as (\ref{1-1}) in the lowest order 
of $g$.\cite{Kajantie,Yaffe}
In this paper, since we deal with the quantum gluonic field described 
by the variables $G$ and $\Sigma$ with ${\ovl {\mib A}}={\mib 0}$, so 
we can evaluate the usual shear viscosity $\eta_C$ 
with higher order of $g$ 
compared with those developed in the previous papers. 
Hereafter, we denote $\eta_C$ as $\eta$ simply because, as a final result in 
this paper, 
the small value of $\eta_C$ can be derived from the viewpoint of 
the weak coupling QCD. 
Thus, it is expected that $\eta$ is also small from Eq.(\ref{1-2}) for the 
quantum gluonic matter.

In order to calculate the shear viscosity, the energy-momentum tensor 
for the pure gluonic field is necessary. 
In the symmetric representation, the energy-momentum tensor 
$T^{\mu}_{\nu}$ is obtained as 
\beq\label{4-8}
& &T^{\mu}_{\nu}=F_a^{\mu\rho}F_{\rho\nu}^a
-\frac{1}{4}\delta_\nu^\mu F_{\rho\sigma}^a F_a^{\rho\sigma} \ . 
\eeq
Then, $T^\mu_\nu$ can be expressed in terms of the color electric and 
color magnetic fields as 
\beq\label{4-9}
& &T_{00}({\mib r})
=\frac{1}{2}({\mib E}^a({\mib r})\cdot{\mib E}^a({\mib r})
+{\mib B}^a({\mib r})\cdot{\mib B}^a({\mib r})) \ , 
\nonumber\\
& &T_{0i}({\mib r})=-
\epsilon_{ijk}E_j^a({\mib r}) B_k^a({\mib r}) \ , \nonumber\\
& &T_{ij}({\mib r})=
-E_i^a({\mib r})E_j^a({\mib r})-B_i^a({\mib r})B_j^a({\mib r})
+\frac{1}{2}\delta_{ij}({\mib E}^a({\mib r})\cdot{\mib E}^a({\mib r})
+{\mib B}^a({\mib r})\cdot{\mib B}^a({\mib r})) \ . \nonumber\\
& &
\eeq

The shear viscosity $\eta(\omega)$ is obtained by taking 
the current $J$ as $T_{xy}=T_{12}$ in Eq.(\ref{4-6}):\cite{Iwasaki}
\beq\label{4-10}
& &\eta(\omega)=
\frac{i}{\omega}\left[
\Pi^{\rm R}(\omega)-\Pi^{\rm R}(0)\right] \ , \nonumber\\
& &\ \ 
\Pi^{\rm R}(\omega)=-i\lim_{\epsilon \rightarrow +0}
\int_0^\infty dt \int d^3{\mib r}
e^{i\omega t-\epsilon t}
\langle\ [\ T_{12}({\mib r},t)\ , \ T_{12}({\mib 0},0)\ ] \ \rangle_{\rm eq} 
\ . 
\eeq
By taking a limit $\omega\rightarrow 0$, the shear viscosity $\eta(0)$ 
has a simple form\cite{Iwasaki} as 
\b\label{4-11}
\eta(0)=-\frac{d}{d\omega}{\rm Im}\ \Pi^{\rm R}(\omega) 
\biggl|_{\omega\rightarrow +0} \ .
\end{equation}
Here, the thermal average $\langle \cdots \rangle_{\rm eq}$ 
can be replaced to the expectation value given in Eq.(\ref{3-21}) 
for $G$ 
and (\ref{2-23b}) for $\Sigma$ 
with $G_k^T$ instead of $G_k$ in our variational 
approach at finite temperature. 
Thus, it is necessary for calculating the shear viscosity to evaluate 
the thermal average 
$\langle[T_{12}({\mib r},t)\ , \ T_{12}({\mib 0},0)]\rangle$ in our framework. 

However, we need the operator at time $t$, namely, $T_{12}({\mib r},t)$. 
In order to derive this operator at $t$, we only need to evaluate 
the operator ${\mib E}^a$ and ${\mib B}^a$ at time $t$ because 
the expression of $T_{12}({\mib r},t)$ has same dependence 
with respect to 
${\mib E}^a$ and ${\mib B}^a$, that is, 
\beq\label{4-12}
T_{12}({\mib r},t)=-E_1^a({\mib r},t)E_2^a({\mib r},t)
-B_1^a({\mib r},t)B_2^a({\mib r},t) \ . 
\eeq
The operators ${\mib E}^a({\mib r},t)$ and ${\mib B}^a({\mib r},t)$ 
are obtained as a result of the time evolution governed by the 
Hamiltonian $\int d^3{\mib x}{\cal H}_0$ with (\ref{2-10}): 
\beq
& &E_i^a({\mib r},t)=e^{iH_0t}E_i^a({\mib r})e^{-iH_0t} \ , \quad
B_i^a({\mib r},t)=e^{iH_0t}B_i^a({\mib r})e^{-iH_0t} \ , 
\label{4-13}\\
& &\qquad
H_0=\int d^3{\mib r}{\cal H}_0=
\int d^3{\mib r} \frac{1}{2}[({\mib E}^a({\mib r}))^2+
({\mib B}^a({\mib r}))^2] \ .
\label{4-14}
\eeq
Here, we can derive 
\beq\label{4-15}
& &[\ H_0 \ , \ E_i^a({\mib r})\ ]
= -i\epsilon_{ijk}\partial_k B_j^a({\mib r})
-\frac{i}{2}g\epsilon_{ijk}f^{abc}(B_j^b({\mib r})A_k^c({\mib r})
+A_k^c({\mib r})B_j^b({\mib r})) \nonumber\\
& &\qquad\qquad\qquad\ \ 
= -i({\mib S}\cdot {\hat {\mib p}})_{ik}
B_k^a({\mib r})+O(g) \ , \nonumber\\ 
& &[\ H_0 \ , \ B_i^a({\mib r})\ ]
=i\epsilon_{ijk}\partial_k E_j^a({\mib r})
+\frac{i}{2}g\epsilon_{ijk}f^{abc}(E_j^b({\mib r})A_k^c({\mib r})
+A_k^c({\mib r})E_j^b({\mib r})) \nonumber\\
& &\qquad\qquad\qquad\ \ 
=i({\mib S}\cdot {\hat {\mib p}})_{ik}E_k^a({\mib r})+O(g) , \\
& &\qquad {\hat {\mib p}}=-i\frac{\partial}{\partial {\mib r}} \ , \nonumber
\eeq
where ${\mib S}$ is defined in (\ref{2-20}). 
Thus, up to the lowest order of $g$ in our quantum gluonic matter, 
we can derive the operators at time $t$ 
as 
\beq\label{4-16}
E_i^a({\mib r},t)&=&
{\hat C}_{ij}^{ab}(t)E_j^b({\mib r})+{\hat D}_{ij}^{ab}(t)B_j^a({\mib r})+O(g) 
\ , \nonumber\\
B_i^a({\mib r},t)&=&
{\hat C}_{ij}^{ab}(t)B_j^b({\mib r})-{\hat D}_{ij}^{ab}(t)E_j^a({\mib r})+O(g) 
\ , \\
{\hat C}_{ij}^{ab}(t)&=&\delta^{ab}[\cos(t({\mib S}\cdot 
{\hat {\mib p}}))]_{ij} \ , 
\quad
{\hat D}_{ij}^{ab}(t)=\delta^{ab}[\sin(t({\mib S}\cdot 
{\hat {\mib p}}))]_{ij} \ . 
\nonumber
\eeq
Thus, we can derive the energy-momentum tensor operator at time $t$ 
in Eq.(\ref{4-12}).

From (\ref{4-10}), we need the 
$\langle[T_{12}({\mib r},t),T_{12}({\mib 0},0)]\rangle$ to estimate the 
shear viscosity in quantum gluonic matter. 
After lengthy but straightforward calculation, we can derive 
the following form up to the lowest order of $g$: 
\beq
& &\langle[\ T_{12}({\mib r},t)\ , \ T_{12}({\mib r}',0) \ ] \ \rangle
\nonumber\\
&=&
{\hat M}_{12ij}^{ab}(t)\cdot 2i\partial_l^{{\mib r}}\delta({\mib r}-{\mib r}')
\cdot
\biggl[\epsilon_{j1l}\Xi_{i2}^{ab}({\mib r},{\mib r}',t)
+\epsilon_{j2l}\Xi_{i1}^{ab}({\mib r},{\mib r}',t)\nonumber\\
& &\qquad\qquad\qquad\qquad\qquad\ \ 
+\epsilon_{i1l}\Xi_{2j}^{ab}({\mib r}',{\mib r},t)
+\epsilon_{i2l}\Xi_{1j}^{ab}({\mib r}',{\mib r},t)
\biggl]
\nonumber\\
& &-
{\hat N}_{12ij}^{ab}(t)\cdot i\partial_l^{{\mib r}}\delta({\mib r}-{\mib r}')
\cdot
\biggl[\epsilon_{j1l}\Upsilon_{i2}^{ab}({\mib r},{\mib r}',t)
+\epsilon_{j2l}\Upsilon_{i1}^{ab}({\mib r},{\mib r}',t)\nonumber\\
& &\qquad\qquad\qquad\qquad\qquad\ \ 
-\epsilon_{i1l}\Upsilon_{2j}^{ab}({\mib r}',{\mib r},t)
-\epsilon_{i2l}\Upsilon_{1j}^{ab}({\mib r}',{\mib r},t)
\biggl]
+O(g^2) \ , 
\label{4-17}
\eeq
where we define 
\beq\label{4-18}
& &{\hat M}_{IJij}^{ab}(t)={\hat C}_{Ii}^{ca}(t){\hat C}_{Jj}^{cb}(t)
+{\hat D}_{Ii}^{ca}(t){\hat D}_{Jj}^{cb}(t) \ , \nonumber\\
& &{\hat N}_{IJij}^{ab}(t)={\hat C}_{Ii}^{ca}(t){\hat D}_{Jj}^{cb}(t)
-{\hat D}_{Ii}^{ca}(t){\hat C}_{Jj}^{cb}(t) \ , \nonumber\\
& &\Xi_{ij}^{ab}({\mib r},{\mib r}',t)
=\epsilon_{imk}\partial_m^{\mib r}(G\Sigma)_{kj}^{ab}({\mib r},{\mib r}',t)
+\epsilon_{jmk}\partial_m^{{\mib r}'}(\Sigma G)_{ik}^{ab}
({\mib r},{\mib r}',t)+O(g^2) \nonumber\\
& &\qquad\qquad\quad\ 
=-({\mib S}\cdot{\hat {\mib p}})_{ik}(G\Sigma)_{kj}^{ab}({\mib r},{\mib r}',t)
-(\Sigma G)_{ik}^{ab}({\mib r},{\mib r}',t)
({\mib S}\cdot {\hat {\lvect {\mib p}}})_{kj}+O(g^2) \nonumber\\
& &\qquad\qquad\quad\ 
=\frac{1}{2}\left[
\langle E_i^a({\mib r})B_j^b({\mib r}')\rangle
+\langle B_i^a({\mib r})E_j^b({\mib r}')\rangle
\right]+O(g^2)\ , \nonumber\\
& &\Upsilon_{ij}^{ab}({\mib r},{\mib r}',t)
=-\frac{1}{4}(G^{-1})_{ij}^{ab}({\mib r},{\mib r}',t)
+\epsilon_{ipm}\epsilon_{jkn}\partial_p^{{\mib r}}\partial_k^{{\mib r}'}
G_{mn}^{ab}({\mib r},{\mib r}',t) \nonumber\\
& &\qquad\qquad\quad\ \ \ \ 
-4(\Sigma G\Sigma)_{ij}^{ab}({\mib r},{\mib r}',t)+O(g^2)\nonumber\\
& &\qquad\qquad\quad\ 
=-\frac{1}{4}(G^{-1})_{ij}^{ab}({\mib r},{\mib r}',t)
+({\mib S}\cdot{\hat {\mib p}})_{im}G_{mn}^{ab}({\mib r},{\mib r}',t)
({\mib S}\cdot{\hat {\lvect {\mib p}}})_{nj}+O(g^2) \nonumber\\
& &\qquad\qquad\quad\ 
=\langle B_i^a({\mib r})B_j^b({\mib r}')\rangle
-\langle E_i^a({\mib r})E_j^b({\mib r}')\rangle
%-\frac{1}{2}g^2f^{acd}f^{bef}\epsilon_{ilm}\epsilon_{jkn}
%G_{lk}^{ce}({\mib r},{\mib r}',t)G_{mn}^{df}({\mib r},{\mib r}',t)
+O(g^2)\ , 
\eeq
where we define 
$f({\mib r},{\mib r}',t){\hat {\lvect {\mib p}}}
=-i\partial f({\mib r},{\mib r}',t)
/\partial {\mib r}$. 
Here, we used the fact that $\Sigma$ is of order of $g$ as is seen 
in Eq.(\ref{2-23}).

\subsection{Lowest order approximation for shear viscosity in 
quantum gluonic matter}

In the lowest order of $g$, ${\hat M}_{12ij}^{ab}$ in Eq.(\ref{4-18}) 
is proportional to $\delta^{ab}$. 
On the other hand, $\Xi_{ij}^{ab}$ is proportional to 
$f^{abc}\omega^c$ because both 
the variables $(G\Sigma)_{kj}^{ab}$ 
and $(\Sigma G)_{ik}^{ab}$ in $\Xi_{ij}^{ab}$ 
are proportional to $f^{abc}\omega^c$ which is derived from 
(\ref{2-23b}) with $G_k^T$ and (\ref{3-21}). 
Thus, in Eq.(\ref{4-17}), the first term vanishes, namely, 
\beq\label{4-19}
& &{\hat M}_{ijkl}^{ab}\Xi_{mn}^{ab}=0 \quad {\rm for\ any}\ i,j,k,l,m\ 
{\rm and}\ n \ .
\eeq

Next, let us consider the second term in (\ref{4-17}). 
At zero temperature, the equation of motion for $G$ can be derived 
from (\ref{2-19}) or (\ref{2-21b}) by 
$\delta \langle H \rangle/\delta G =0$ which leads to 
\beq\label{4-20}
-\frac{1}{8}G^{-2}+\frac{1}{2}K=0
\eeq
in the lowest order of $g$. Thus, the solution of $G$ is written as 
\beq\label{4-21}
G=\frac{1}{2\sqrt{K}}=\frac{1}{2{\mib S}\cdot {\hat {\mib p}}} \ , 
\eeq
where $K=({\mib S}\cdot{\hat {\mib p}})^2$ with ${\ovl {\mib A}}={\mib 0}$ 
in (\ref{2-20}). 
Using the above fact, from (\ref{4-18}), the following is derived up to 
the lowest order of $g$: 
\beq\label{4-22}
\Upsilon_{ij}^{ab}=-\left[\frac{1}{4}G^{-1}-
({\mib S}\cdot{\hat {\mib p}})G({\mib S}\cdot{\hat {\mib p}})\right]_{ij}^{ab}
=0 \ .
\eeq
As a result, the shear viscosity for the pure quantum gluonic matter 
is zero at zero temperature up to the order of $g^1$: 
\beq\label{4-23}
\eta(\omega)=0 
\eeq
up to the order of $g$ at zero temperature.

At finite temperature, we have the solution for $G$ and $\Sigma$ in 
Eq.(\ref{3-21}) and 
(\ref{2-23b}) with $G_k^T$ instead of $G_k$. 
Thus, we obtain $\Upsilon$ in the lowest order approximation as 
\beq\label{4-24}
\Upsilon_{ij}^{ab}({\mib r},{\mib r}',t)
&=&\delta^{ab}\int \frac{d^3{\mib k}}{(2\pi)^3}
e^{i{\mib k}\cdot ({\mib r}-{\mib r}')}\cdot
2|{\mib k}|\cdot\frac{n_k(n_k+1)}{2n_k+1}
\left(\delta_{ij}-\frac{k_ik_j}{|{\mib k}|^2}\right) \nonumber\\
&=&\delta^{ab}\int \frac{d^3{\mib k}}{(2\pi)^3}
e^{i{\mib k}\cdot ({\mib r}-{\mib r}')}\cdot
\frac{|{\mib k}|}{\sinh \left(\frac{E_k}{T}\right)}
\left(\delta_{ij}-\frac{k_ik_j}{|{\mib k}|^2}\right) 
\eeq
From Eq.(\ref{4-10}), the complex admittance for the shear viscosity 
$\eta(\omega)$ is 
written as 
\beq\label{4-25}
\Pi^{\rm R}(\omega)&=&-i\lim_{\epsilon \rightarrow +0}\int_{0}^{\infty} dt
\int d^{3}{\mib r}\ e^{i\omega t-\epsilon t}\langle
[\ T_{12}({\mib r},t)\ , \ T_{12}({\mib 0},0)\ ]\rangle \nonumber\\
&=&
i\lim_{\epsilon \rightarrow +0}\int_{0}^{\infty} dt
\int d^{3}{\mib r}\ e^{i\omega t-\epsilon t}
{\hat N}_{12ij}^{ab}(t)(i\partial_l \delta^3({\mib r})) \nonumber\\
& &\qquad\quad
\times
[ \epsilon_{j1l}\Upsilon_{i2}^{ab}({\mib r},{\mib 0},t)
+\epsilon_{j2l}\Upsilon_{i1}^{ab}({\mib r},{\mib 0},t)
-\epsilon_{i1l}\Upsilon_{2j}^{ab}({\mib 0},{\mib r},t)
-\epsilon_{i2l}\Upsilon_{1j}^{ab}({\mib 0},{\mib r},t)] \nonumber\\
&=&i\lim_{\epsilon \rightarrow +0}\int_{0}^{\infty} dt
\int d^{3}{\mib r}\ e^{i\omega t-\epsilon t}
{\hat N}_{12ij}^{ab}(t)\bigl[\delta^3({\mib r}) \nonumber\\
& &\quad
\times\{\epsilon_{j1l}\Upsilon_{l,i2}^{ab}({\mib r},{\mib 0},t)
+\epsilon_{j2l}\Upsilon_{l,i1}^{ab}({\mib r},{\mib 0},t)
+\epsilon_{i1l}\Upsilon_{l,2j}^{ab}({\mib 0},{\mib r},t)
+\epsilon_{i2l}\Upsilon_{l,1j}^{ab}({\mib 0},{\mib r},t)\}\bigl] \ , 
\nonumber\\
& &
\eeq
where the integration by part has been carried out from the second 
line to the third line and we define $\Upsilon_{l,ij}({\mib r},{\mib r}',t)$ 
as 
\beq\label{4-26}
\Upsilon_{l,ij}({\mib r},{\mib r}',t)
=\delta^{ab}\int \frac{d^3{\mib k}}{(2\pi)^3}
e^{i{\mib k}\cdot ({\mib r}-{\mib r}')}
\frac{k_l|{\mib k}|}{\sinh\left(\frac{E_k}{T}\right)}
\left(\delta_{ij}-\frac{k_ik_j}{|{\mib k}|^2}\right) \ .
\eeq
Here, up to the lowest order of $g$ in this quantum gluonic matter, 
the operator ${\hat N}_{12ij}^{ab}(t)$ is written in Eq.(\ref{4-18})
with (\ref{4-16}). Thus, we can expand ${\hat N}_{12ij}^{ab}(t)$ 
as 
\beq\label{4-27}
{\hat N}_{12ij}^{ab}(t)
&=&\delta^{ab}\sum_{n=1}\left[
f(2n-1)t^{2n-1}({\mib S}\cdot{\hat {\mib p}})^{2n-1}\right]_{12ij}\nonumber\\
&=&\delta^{ab}\left[
\delta_{1i}t({\mib S}\cdot{\hat {\mib p}})_{2j}-
\delta_{2j}t({\mib S}\cdot{\hat {\mib p}})_{1i} +O({\hat {\mib p}^3})\right]
\ ,
\eeq
where $f(x)$'s are expansion matrices with 
numerical factors for this expansion and 
${\hat {\mib p}}=-i\partial/\partial {\mib r}$. 
Thus, 
\beq\label{4-28}
\int d^3{\mib r} {\hat N}_{12ij}^{ab}(t)\left[\delta^3({\mib r})
\Upsilon_{l,km}^{ab}({\mib r},{\mib 0},t)\right]=0
\eeq 
for any $i$, $j$, $k$ and $m$ because the integrand 
is the total derivative with respect to ${\mib r}$. 
Thus, we can obtain the relation $\Pi^{\rm R}(\omega)=0$. 
Finally, from Eqs.(\ref{4-25}) and (\ref{4-28}), 
we conclude that the shear viscosity $\eta(\omega)$ in Eq.(\ref{4-10}) 
is as follows 
\beq\label{4-29}
\eta(\omega)=0
\eeq
up to the lowest order of $g$ in the quantum gluonic matter. 
Here, the terms of the next order of ${\hat C}_{ij}^{ab}(t)$ 
and ${\hat D}_{ij}^{ab}(t)$ in ${\hat N}_{12ij}^{ab}(t)$ 
include the coupling constant $g$ without the spatial 
derivative. Thus, Eq.(\ref{4-28}) is not satisfied in the order of $g$. 
Therefore, the result (\ref{4-29}) is valid up to the order of 
$g^0$ while (\ref{4-23}) is valid up to the order of $g^1$ at zero 
temperature because the mechanism to vanish the value of shear viscosity is 
different.

%%%%%%%%%%%%%%%%%%%%%%%%%%%%%%%%%%%%%%%%%%%%%%%%%%%%%%%%%%%%%%%%%%%%%%%

%%%%%%%%%%%%%%%%%%%%%%%%%%%%%%%%%%%%%%%%%%%%%%%%%%%%%%%%%%%%%%%%%%%%%%%%%

%%%%%%%%%%%%%%%%%%%%%%%%%%%%%%%%%%%%%%%%%%%%%%%%%%%%%%%%%%%%%%%%%%%%%%
%\begin{figure}[t]
%\begin{center}
%\includegraphics[height=5.3cm]{fig3.eps}
%\caption{The sigma and pion masses are shown as functions of temperature 
%$T$ in (a). The order parameter of the chiral phase transition 
%$\varphi_0$ is depicted in (b). 
%The cutoff parameter $\Lambda$ is taken as 650 MeV.
%}
%\label{fig:4-1}
%\end{center}
%\end{figure}
%%%%%%%%%%%%%%%%%%%%%%%%%%%%%%%%%%%%%%%%%%%%%%%%%%%%%%%%%%%%%%%%%%%%%%%%

\section{Summary and concluding remarks}

In this paper, the time-dependent variational method for the pure Yang-Mills 
gauge theory is formulated in the functional Schr\"odinger picture with the 
Gaussian trial wave functional. In this variational method, the 
classical mean fields and the quantum fluctuations around them are 
treated self-consistently and both degrees of freedom are coupled each other. 
Further, the equations of motion for the quantum fluctuations around the 
mean fields were 
reformulated in a form of the Liouville-von Neumann equation for the 
reduced density matrix which was introduced in the Hartree-Bogoliubov 
approximation developed in the many-body problems for boson systems.

This variational method developed in this paper was applied to the 
pure quantum gluonic matter system in order to evaluate the shear viscosity, 
which is one of the transport coefficients of the gluonic matter 
in the system with the color $su(3)$ symmetry, namely, the QCD without quarks. 
As a result, it was shown that there is 
no contribution of the quantum gluons to the shear 
viscosity in the pure gluonic matter up to the lowest order of the QCD 
coupling $g$ at finite temperature. 
Namely, up to the order of $g^0$, the contribution of the quantum gluons to 
the shear viscosity is nothing. 
At zero temperature, adding to the order of $g^0$, 
there is no contribution up to the order of 
$g$ due to the equations of motion. 
Thus, for small $g$, namely, from the viewpoint of the weak coupling 
QCD,\cite{Asakawa} 
the shear viscosity in quantum gluonic matter 
may be small because 
the quantum gluons contribute to the shear viscosity from the order of 
$g$ or higher at finite temperature.

Recently, Matsui and Matsuo give the transport equations for the Wigner 
distribution function and the anomalous distribution function\cite{MM} 
which are 
coupled each other to determine the dynamics of the meson fields and the 
fluctuations around them in the linear sigma model, as is similar to 
our formalism. 
In order to compare the theoretical analysis with the experimental 
results, the information of the gluon distribution function may be 
necessary as was discussed in Ref.\citen{MM} in the case of the 
linear sigma model. 
It is one of important further problems to investigate the gluon 
distribution function governed by the transport equation derived by 
this formalism such as the 
extended Boltzmann equations\cite{MM}.

It may be interesting to investigate 
the higher order contribution to the shear viscosity 
because the approximation used in this paper corresponds to 
the Hartree-Bogoliubov like approximation. 
The random phase approximation (RPA) is missing in the treatment in 
this paper. 
It may be necessary to extend our treatment to including the RPA like 
modes as was developed for the linear sigma model.\cite{TM09}
Further, it may be also interesting to investigate 
the behavior of other transport coefficients in this framework developed in 
this paper. 
They are future problems.

\section*{Acknowledgements} 
%\acknowledgement 
The authors would like to express their sincere thanks to Professors 
M. Iwasaki and K. Iida, Drs. T. Saito and K. Ishiguro 
and the members of Many-Body Theory 
Group of Kochi University for discussing the subjects in this 
paper and 
giving them valuable comments. 
One of the authors (Y.T.) also would like to express his sincere thanks to 
Professor 
Dominique 
Vautherin for the collaboration and giving him 
the suggestion for this work developed in this paper. 
He also thanks to Professor\break
T. Matsui and Dr. M. Matsuo for informing 
him about the formalism of the transport equations for the usual and 
the anomalous distribution functions derived by the similar formalism to 
theirs in the linear sigma model. 
He is partially supported by the Grants-in-Aid of the Scientific Research 
No.18540278 from the Ministry of Education, Culture, Sports, Science and 
Technology in Japan.

%\appendix
%\section{}

\end{document}